\documentclass[a4paper,11pt]{article}
\usepackage{jcappub}

\usepackage[colorlinks=true,urlcolor=blue,linkcolor=blue]{hyperref}
\usepackage{graphicx}
\usepackage{epstopdf}
\usepackage{color}
\usepackage{epsfig}
\relax

\def\be{\begin{equation}}
\def\ee{\end{equation}}
\def\bs{\begin{subequations}}
\def\es{\end{subequations}}

\def\be{\begin{equation}}
\def\ee{\end{equation}}
\def\bs{\begin{subequations}}
\def\es{\end{subequations}}
\newcommand{\een}{\end{subequations}}
\newcommand{\ben}{\begin{subequations}}
\newcommand{\beq}{\begin{eqalignno}}
\newcommand{\eeq}{\end{eqalignno}}

\newcommand{\apj}{ApJ}
\newcommand{\mnras}{MNRAS}
\newcommand{\prd}{PhysRevD}
\newcommand{\jcap}{JCAP}
\newcommand{\aj}{AJ}

\newcommand{\aap}{A\& A}
\newcommand{\physrep}{PhysRep}

\newcommand\fverb{\setbox\pippobox=\hbox\bgroup\verb}
\newcommand\fverbdo{\egroup\medskip\noindent%
                        \fbox{\unhbox\pippobox}\ }
\newcommand\fverbit{\egroup\item[\fbox{\unhbox\pippobox}]}
\newbox\pippobox

\title{Turnaround overdensity as a cosmological observable: the case for a local measurement of $\mathbf{\Lambda}$}

\author[a,c]{D. Tanoglidis}
\author[a,b,c]{V. Pavlidou}
\author[a,b]{T.~N. Tomaras}
\affiliation[a]{Department of Physics, University of Crete, P.O. Box 2208, Heraklion, 70013 Greece} 
\affiliation[b]{Institute for Theoretical $\&$ Computational Physics, University of Crete, P.O. Box 2208, Heraklion, 70013 Greece}
\affiliation[c]{IESL, Foundation for Research and Technology - Hellas, P.O. Box 1527, Heraklion, 71110 Greece} 
\emailAdd{dtanogl@physics.uoc.gr}
\emailAdd{pavlidou@physics.uoc.gr}
\emailAdd{tomaras@physics.uoc.gr}

\abstract{We demonstrate that, in the context of the $\Lambda$CDM model, it is in principle possible to measure the value of the cosmological constant by tracing, across cosmic time, the evolution of the turnaround radius of cosmic structures. The novelty of the presented method is that it is \emph{local}, in the sense that it uses the effect of the cosmological constant on the relatively short scales of cosmic structures and not on the dynamics of the Universe at its largest scales. In this way, it can provide an important consistency check for the standard cosmological model and can give signs of new physics, beyond $\Lambda$CDM.}
\begin{document}

\maketitle
\flushbottom

\section{Introduction}
\setcounter{equation}{0}

The existence of a dark energy component in the matter-energy content of the Universe is now supported by a diverse set of observations (e.g., \cite{Amendola, Mort et al}). All current observations are consistent with the simplest possible candidate, a cosmological constant $\Lambda$, whose density remains constant in time (e.g., \cite{Amendola, Mort et al} and \cite{Planck,Planck2} ). The most explored alternative is a dark energy component with an equation of state parameter $w \neq -1$ (e.g., \cite{Phantom energy, Caroll} and references therein). However,  more exotic alternatives are also pursued, including living in a giant cosmic void that leads to an apparent global acceleration \cite{Void1,Void2,Void3,Void4}, or deviations from General Relativity at the largest scales (e.g., models with screening mechanisms, such as the galileon model) \cite{ModGrav1,ModGrav2,ModGrav3,Screen}.

Although observations pointing towards the existence of dark energy vary in nature, they all share a common feature: they all consider the Universe at its largest scales, by measuring, for example, its --apparently-- accelerated expansion through observations of distant supernovae (e.g., \cite{Riess, Perlmutter, Schmidt}) or by tracing the discrepancy between the measured matter density of the Universe (from measurements of BAO's, e.g., \cite{BAOS1,BAOS2} and references therein) and the energy density needed for the Universe to be flat as a whole, as CMB observations indicate (e.g., \cite{Planck}). It may still thus be possible that the above are just manifestations of our ignorance of physics at the largest scales. For this reason, a ``local" test probing the existence of dark energy would be a powerful complement to our observational cosmology tool arsenal, in order to check for consistency results obtained using the above methods.

A dark energy component in the form of a cosmological constant  has a prominent effect on the process of structure formation: acting ``anti-gravitationally", it halts structure growth (\cite{Busha1,Busha2} and \cite{PavTom,myRoy}). In such a cosmology, a clear prediction can be made of the maximum turnaround radius --the non-expanding shell furthest away from the center of a bound structure-- a cosmic structure can have. In \cite{PavTom} it is shown that this maximum value for a structure of mass $M$ is equal to: 
\begin{equation} \label{turadius}
R_{\mbox{\scriptsize{ta,max}}}=\left(\frac{3GM}{\Lambda c^2} \right)^{1/3},
\end{equation}
where $G$ is Newton's gravitational constant and $c$ is the speed of light. This requirement can be used to construct cosmological test, which is to find \emph{non-expanding} structures with radii violating the bound \eqref{turadius}.

  Even though this test is extremely powerful and robust, in its simplest form it is not flexible enough to provide constraints on the cosmological parameters. Rather,  if the Universe is indeed $\Lambda$CDM, it can only provide a null result; if not, it can point to the need of a different cosmological model.  In the present work, we overcome this shortcoming by extending the $\Lambda$CDM predictions of the turnaround radius of structures beyond that infinite time in the future (corresponding to the maximum value, eq. \eqref{turadius}): rather, we present a prediction for every cosmic epoch and for any possible combination of $ \Omega_{\mbox{\scriptsize{m,0}}}$ and $\Omega_{\mbox{\scriptsize{$\Lambda$,0}}}$.

Our aim is to demonstrate that it is \emph{in principle} possible to use measurements of turnaround masses and radii of structures, at different epochs, in order to determine the values of $ \Omega_{\mbox{\scriptsize{m,0}}}$ and $\Omega_{\mbox{\scriptsize{$\Lambda$,0}}}$. Particularly, we show that it is possible to construct a new test that can prove that a non-zero cosmological constant exists, using \emph{local} physics --not concerning the Universe as a whole. Such a proof --if verified-- would present an extreme challenge for more exotic alternatives to dark energy, such as those described above. On the other hand, a difficulty to firmly detect a $\Lambda \neq 0$ using this method, would constitute a strong indication of physics beyond $\Lambda$CDM. In both cases, we would have an extremely important result. This work gives a ``proof of principle" for the feasibility of such a test.
  
Our paper is organized as follows: In section \ref{Model}, we present a simple model to describe the time evolution of the turnaround radius and its dependence on the cosmological model. In section \ref{Dependence}, we demonstrate the different evolution of turnaround radii for models with different values of $ \Omega_{\mbox{\scriptsize{m,0}}}$ and $\Omega_{\mbox{\scriptsize{$\Lambda$,0}}}$. In section \ref{measurement}, we present how these results can be used to construct a way to measure $\Lambda$ locally and we give a general discussion of our work in section \ref{Discussion}.

\section{The model} \label{Model}

\subsection{Time evolution of the turnaround radius: general considerations} \label{timeevol}

We start by considering the evolution of the turnaround radius of a cosmic structure. Let $\delta_{\mbox{\scriptsize{ta}}}(a)$ be the overdensity of a turnaround structure at cosmic epoch $a$, as obtained from the spherical collapse model (see also $\S$ \ref{overdensity}). Then, from its definition \cite{Padma}:
\begin{equation} \label{overd}
\delta_{\mbox{\scriptsize{ta}}}(a) \equiv \frac{\rho_{\mbox{\scriptsize{ta}}}(a)-\rho_{\mbox{\scriptsize{m}}}(a)}{\rho_{\mbox{\scriptsize{m}}}(a)},
\end{equation}
where $\rho_{\mbox{\scriptsize{ta}}}(a)$ is the density of the turnaround structure at turnaround and at cosmic epoch $a$, and $\rho_{\mbox{\scriptsize{m}}}(a)$ is the mean matter density of the Universe at the same cosmic epoch. If we further assume that the turnaround structure is a sphere of constant density with total mass $M$, then: 
\begin{equation} \label{density}
\rho_{\mbox{\scriptsize{ta}}}(a) = \frac{M}{\frac{4}{3}\pi R_{\mbox{\scriptsize{ta}}}^3(a)}
\end{equation}
with $R_{\mbox{\scriptsize{ta}}}$ being the turnaround radius. From eqs. \eqref{overd} and \eqref{density}, and using that ${\rho}_{\mbox{\scriptsize{m}}}={\rho}_{\mbox{\scriptsize{m,0}}}/a^3 $, we get the turnaround radius as a function of cosmic epoch $a$:
\begin{equation} \label{radius1}
R_{\mbox{\scriptsize{ta}}}(a)=\left[\frac{3}{4\pi (1+\delta_{\mbox{\scriptsize{ta}}}(a)){\rho}_{\mbox{\scriptsize{m,0}}}}\right]^{1/3} a\,\, {M}^{1/3},
\end{equation}
with ${\rho}_{\mbox{\scriptsize{m,0}}}$ the mean matter density of the Universe today.

In a recent work \cite{myRoy}, we have shown that in $\Lambda$CDM a special mass scale exists, which separates structures with qualitatively different cosmological evolution. We have called this mass scale the \emph{transitional} mass scale, which we have calculated to be:
\begin{equation} \label{transitmass}
M_{\mbox{\scriptsize{transitional}}} \simeq 10^{13} M_\odot.
\end{equation}
A convenient normalization of eq. \eqref{radius1} would be one that makes use of the importance of the transitional mass scale. For this reason, we define:
\begin{equation} \label{mstar}
M^\ast \equiv 10^{13} M_\odot,
\end{equation}
which is the order of magnitude of the transitional mass scale. We also define a related length scale, as the radius of the sphere which has mass $M^\ast$ and density equal to the current critical density of the Universe, ${\rho}_{\mbox{\scriptsize{c,0}}}$:
\begin{equation} \label{rstar}
R^\ast \equiv \left(\frac{3 M^\ast}{4 \pi {\rho}_{\mbox{\scriptsize{c,0}}} } \right)^{1/3} \cong 2.05 \,h^{-2/3}\,\, \mbox{Mpc},
\end{equation}
$h$ being the dimensionless Hubble parameter. Expressing radii and masses in terms of $R^\ast$ and $M^\ast$, respectively, eq. \eqref{radius1} becomes:
\begin{equation} \label{Radius2}
\frac{R_{\mbox{\scriptsize{ta}}}(a)}{R^\ast}=\left[(1+\delta_{\mbox{\scriptsize{ta}}}(a)){\Omega}_{\mbox{\scriptsize{m,0}}} \right]^{-1/3} a \left(\frac{M}{M^\ast}\right)^{1/3}, 
\end{equation}
where ${\Omega}_{\mbox{\scriptsize{m,0}}} \equiv  {\rho}_{\mbox{\scriptsize{m,0}}}/{\rho}_{\mbox{\scriptsize{c,0}}}$.

We can isolate the part which gives the time evolution of the turnaround radius that does not depend on the mass of the structure, by taking logarithms on both parts of eq. \eqref{Radius2}:
\begin{equation} \label{logplot}
\log\left(\frac{R_{\mbox{\scriptsize{ta}}}(a)}{R^\ast} \right)=\frac{1}{3}\log\left(\frac{M}{M^\ast}\right)+\log\left(\left[(1+\delta_{\mbox{\scriptsize{ta}}}(a)){\Omega}_{\mbox{\scriptsize{m,0}}} \right]^{-1/3} a \right).
\end{equation}
The mentioned time evolution has been isolated in the second logarithm of the R.H.S. of eq. \eqref{logplot}. We define this part as:
\begin{equation} \label{Idef}
I(a) \equiv \log\left(\left[(1+\delta_{\mbox{\scriptsize{ta}}}(a)){\Omega}_{\mbox{\scriptsize{m,0}}} \right]^{-1/3} a \right).
\end{equation}
With this definition:
\begin{equation} \label{Idef2}
\log\left(\frac{R_{\mbox{\scriptsize{ta}}}(a)}{R^\ast} \right)=I(a) +\frac{1}{3}\log\left(\frac{M}{M^\ast}\right).
\end{equation}
For a theoretical calculation of $I(a)$ we need to know $\delta_{\mbox{\scriptsize{ta}}}(a)$, i.e. how the turnaround overdensity evolves with time for different cosmologies. In the following subsection, we show how we can calculate this overdensity for different combinations of $\Omega_{\mbox{\scriptsize{m,0}}}$ and $\Omega_{\mbox{\scriptsize{$\Lambda$,0}}}$, in the context of the simple spherical top-hat model.

\subsection{Calculation of the turnaround overdensity} \label{overdensity}

The turnaround overdensity is given by \cite{PikaiFi}:
\begin{equation} \label{delta}
\delta_{\scriptsize{\mbox{ta}}}(a)=\left(\frac{a}{a_{\scriptsize{\mbox{p,ta}}}(a)}\right)^3-1,
\end{equation}
where $a_{\scriptsize{\mbox{p,ta}}}(a)$ is the scale factor of a spherical perturbation/overdensity that turns around at time $a$, $a$ being the scale factor of the Universe. The evolution of a spherical matter overdensity in a background universe, not necessarily flat, with matter and cosmological constant, is dictated by the equation (which is obtained by dividing the Friedmann equation for the perturbation with the Friedmann equation for the background Universe):
\begin{equation} \label{evoleq}
\left( \frac{d a_{\scriptsize{\mbox{p}}}}{da}\right)^2 = \frac{a}{a_{\scriptsize{\mbox{p}}}} \frac{\omega  a_{\scriptsize{\mbox{p}}}^3-\kappa a_{\scriptsize{\mbox{p}}}+1}{\omega a^3 +\xi a +1},
\end{equation}
together with the initial condition $a_{\scriptsize{\mbox{p}}}(0)=0$. Also we have the definitions:
\begin{equation}
\omega \equiv \frac{\Omega_{\scriptsize{\Lambda,0}}}{\Omega_{\scriptsize{\mbox{m},0}}}, \qquad \xi \equiv \frac{1-\Omega_{\scriptsize{\mbox{m,0}}}-\Omega_{\scriptsize{\Lambda,0}}}{\Omega_{\scriptsize{\mbox{m,0}}}} = \frac{1}{\Omega_{\scriptsize{\mbox{m,0}}}}-1-\omega
\end{equation}
while, for an overdensity, $\kappa$ is a positive constant that indicates its magnitude.

 At turnaround $d a_{\scriptsize{\mbox{p}}} / da = 0$ (the perturbation reaches maximum size and starts collapsing). This requires that there is a real and positive solution to the equation: 
\begin{equation}
\omega  a_{\scriptsize{\mbox{p,ta}}}^3-\kappa a_{\scriptsize{\mbox{p,ta}}}+1=0.
\end{equation}
There is a minimum value of the magnitude of the overdensity, $\kappa$, for which this happens and can be obtained by expressing $\kappa$ as:
\begin{equation} \label{kappa}
\kappa = \frac{\omega  a_{\scriptsize{\mbox{p,ta}}}^3+1}{a_{\scriptsize{\mbox{p,ta}}}}
\end{equation}
and finding the $a_{\scriptsize{\mbox{p,ta}}}$ which minimizes it, by setting $d \kappa / d a_{\scriptsize{\mbox{p,ta}}} = 0$. This gives the minimum value of $\kappa$ for turnaround and collapse, and the corresponding maximum value of the the turnaround turnaround radius: 
\begin{equation}
\kappa_{\scriptsize{\mbox{min,coll}}}=\frac{3\omega^{1/3}}{2^{2/3}}, \qquad a_{\scriptsize{\mbox{p,ta,max}}}=(2 \omega)^{-1/3}.
\end{equation}
Note that using the second equation of the above, the definition of $\omega$, and also eqs. \eqref{delta} and \eqref{radius1}, we get the physical maximum turnaround radius, eq. \eqref{turadius}. In what follows, we are working with perturbations with $\kappa > \kappa_{\scriptsize{\mbox{min,coll}}}$ (or, equivalently, with $a<a_{\scriptsize{\mbox{p,ta,max}}})$ for which the relation \eqref{kappa} holds.

Taking the positive square root of \eqref{evoleq}, separating variables and integrating (using also the initial condition) we get:
\begin{equation} \label{bigeq}
\int_0^{a_{\scriptsize{\mbox{p}}}} \frac{\sqrt{y}}{\sqrt{\omega y^3 - \kappa y +1} \quad dy} = \int_0^a \frac{\sqrt{x}}{\sqrt{\omega x^3 + \xi x +1}} \quad dx
\end{equation}
Making the change of variables $u = \frac{y}{a_{\scriptsize{\mbox{p,ta}}}}$, defining $r= \frac{a_{\scriptsize{\mbox{p}}}}{a_{\scriptsize{\mbox{p,ta}}}}$ and using eq. \eqref{kappa} in the L.H.S. integral and  also performing a change of variables $y=x/a$ in the R.H.S. integral, we rewrite the above equation as:
\begin{equation}
\int_0^r \frac{\sqrt{u}}{\sqrt{(1-u)\left[\frac{1+\delta_{\scriptsize{\mbox{ta}}}(a)}{\omega a^3} - u(u+1) \right]}} \quad du = \omega^{1/2} \int_0^{1} \frac{\sqrt{y}}{\sqrt{\omega y^3 + \xi \frac{y}{a^2} +\frac{1}{a^3}}} \quad dy.
\end{equation}
Noting that at turnaround $r=1$, we finally get: 
\begin{equation}\label{obtdelt}
\int_0^1 \frac{\sqrt{u}}{\sqrt{(1-u)\left[\frac{1+\delta_{\scriptsize{\mbox{ta}}}(a)}{\omega a^3} - u(u+1) \right]}} \quad du = \omega^{1/2} \int_0^{1} \frac{\sqrt{y}}{\sqrt{\omega y^3 + \xi \frac{y}{a^2} +\frac{1}{a^3}}} \quad dy.
\end{equation}
For a given $a$, we can solve numerically eq. \eqref{obtdelt} to obtain $\delta_{\scriptsize{\mbox{ta}}}(a)$.

The case of a Universe without cosmological constant (only with matter) gives a compact expression for $a_{\scriptsize{\mbox{p,ta}}}(a)$. Setting $\omega=0$ in eq. \eqref{evoleq} and in the definition of $\xi$, and working as before we get:
\begin{equation} \label{matteronly}
a_{\scriptsize{\mbox{p,ta}}}(a)= \left[\frac{2}{\pi} \int_0^{a} \frac{\sqrt{x}}{\sqrt{1+\left(\frac{1-\Omega_{\scriptsize{\mbox{m,0}}}}{\Omega_{\scriptsize{\mbox{m,0}}}} \right)x}} \quad dx  \right]^{2/3}
\end{equation}
For $\Omega_{\scriptsize{\mbox{m,0}}}=1$, from \eqref{matteronly} and \eqref{delta} we get:
\begin{equation}
\delta_{\scriptsize{\mbox{ta}}}(a) \cong 4.55,
\end{equation}
constant for every cosmic epoch, which is a well-known result in the literature (e.g., \cite{Padma,Peebles}).

For a more general discussion about spherical collapse in a $\Lambda$CDM Universe see, for example, \cite{PikaiFi,Eke,Lokas,Barrow}.

\section{Dependence of the time evolution of the turnaround radius on $\Omega_{\mbox{\scriptsize{m,0}}}$ and $\Omega_{\mbox{\scriptsize{$\Lambda$,0}}}$} \label{Dependence}

In \S \ref{timeevol} we reduced the problem of describing the time evolution of the turnaround radius of cosmic structures to the calculation of the function $I(a)$, eq. \eqref{Idef}. The exact form of this function depends on cosmology explicitly through the value of $\Omega_{\mbox{\scriptsize{m,0}}}$ and also through the turnaround overdensity,  $\delta_{\mbox{\scriptsize{ta}}}(a)$, which is different for different combinations of $\Omega_{\mbox{\scriptsize{m,0}}}$ and $\Omega_{\mbox{\scriptsize{$\Lambda$,0}}}$. In \S \ref{overdensity} we showed how to use the spherical collapse model in order to calculate the turnaround overdensity for a general combination of these parameters.

In the present section, we use these results to demonstrate how the values of the present matter and cosmological constant density parameters are imprinted in the evolution history of the turnaround radius. For this reason we plot the theoretical predictions for $I(a)$ for different cosmologies. The results presented here show that the existence of a cosmological constant in the Universe has a profound local effect that --in principle-- can be measured. In the following section we will build upon this, in order to propose a way for a local measurement of the cosmological constant density.

\subsection{Constant matter density}

We start by considering models with different values for the cosmological constant energy density, $\Omega_{\mbox{\scriptsize{$\Lambda$,0}}}$, but with the same value for the matter density, $\Omega_{\mbox{\scriptsize{m,0}}}=0.30$ (roughly the currently accepted value). For the $\Omega_{\mbox{\scriptsize{$\Lambda$,0}}}$ we have chosen values in the range $[0.00,1.50]$. Our aim is to show that, following the evolution of turnaround radii of structures, we are in principle able to distinguish between different values of $\Omega_{\mbox{\scriptsize{$\Lambda$,0}}}$; especially between models with and without a cosmological constant.

\begin{figure} 
\centering
\includegraphics[scale=0.6, clip]{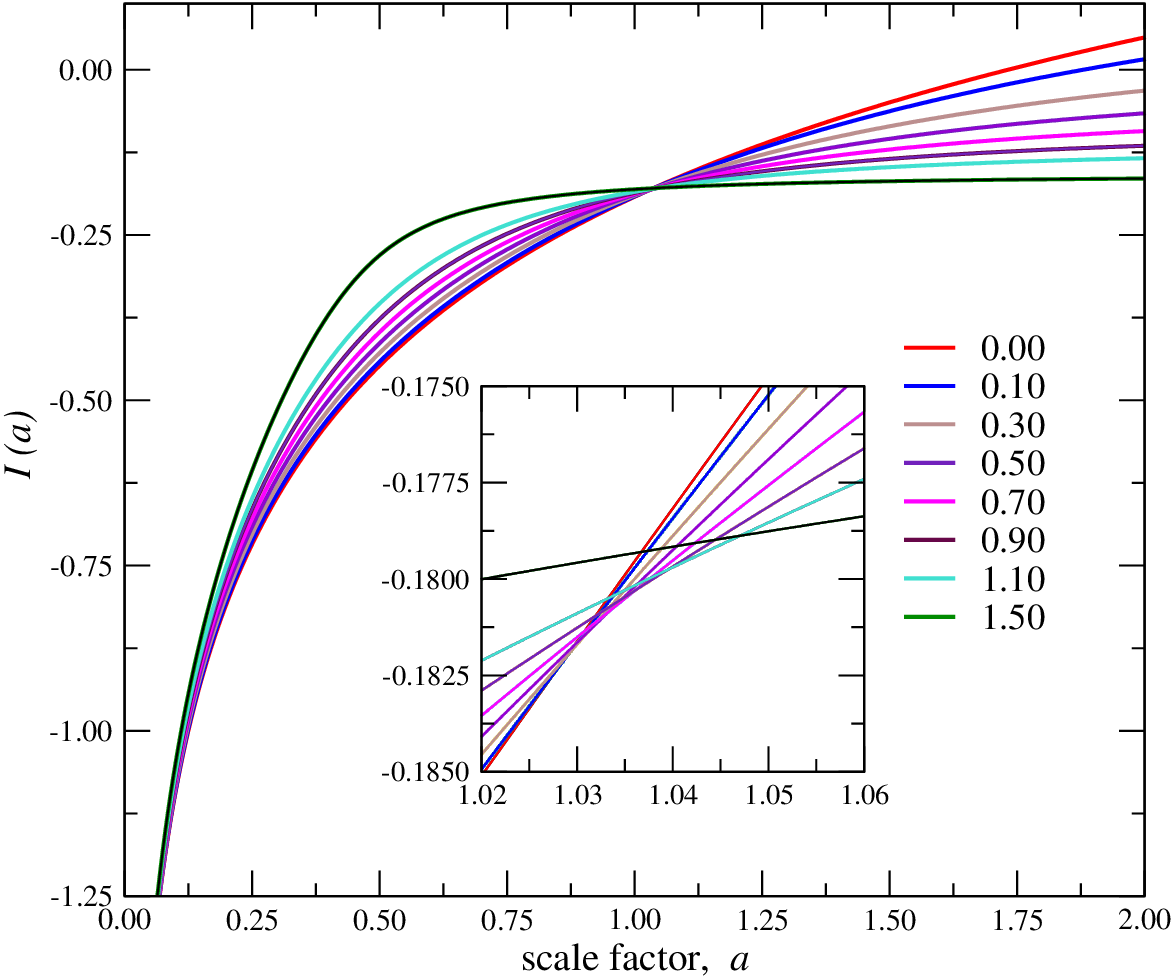} 
\caption{$I(a)$ as a function of $a$, for constant $\Omega_{\mbox{\scriptsize{m,0}}}=0.30$, and different values of $\Omega_{\mbox{\scriptsize{$\Lambda$,0}}}$, in a range from $\Omega_{\mbox{\scriptsize{$\Lambda$,0}}}=0.00$ to $\Omega_{\mbox{\scriptsize{$\Lambda$,0}}}=1.50$. The inset figure presents $I(a)$ for the same parameters as in the greater figure, but in a shorter range in $a$, around the point where all graphs seem to intersect in the greater figure.}
\label{fig: figure1}
\end{figure}

In figure \ref{fig: figure1} we plot $I(a)$, eq. \eqref{Idef}, as a function of $a$, from $a \sim 0$ to $a=2.00$ and for the values of cosmological parameters mentioned above. In an inset figure, we present $I(a)$ for the same sets of parameters, but in a narrower range in $a$, around the point where all graphs of the bigger plot seem to intersect. From this figure we can see how different values of $\Omega_{\mbox{\scriptsize{$\Lambda$,0}}}$ predict different evolution histories, $I(a)$. Let us discuss the main features of this plot.

From the plot it can be inferred that for models with high value for $\Omega_{\mbox{\scriptsize{$\Lambda$,0}}}$, $I(a)$ approaches a constant value. Indeed, as we discuss below, this is also true for any model with a non-zero cosmological constant, even it is not clear in this plot, since in models with a smaller value for $\Omega_{\mbox{\scriptsize{$\Lambda$,0}}}$, $I(a)$ approaches its ultimate value later. Also, we can see that ultimate value of $I(a)$, $I_{\mbox{\scriptsize{ult}}}=I(a \to \infty)$, is smaller, the higher the value of  $\Omega_{\mbox{\scriptsize{$\Lambda$,0}}}$ is. This behaviour can easily be explained,  using eq. \ref{turadius} for the ultimate / maximum value for the turnaround radius. Expressing and masses and radii in terms of $M^\ast$ and $R^\ast$, (eqs.  \eqref{mstar} and \eqref{rstar}) as in \S \ref{timeevol}, and taking logarithms in both sides, we get:
\begin{equation}
\log\left(\frac{R_{\mbox{\scriptsize{ta,max}}}}{R^\ast} \right)=\frac{1}{3}\log \left(\frac{4\pi G \rho_{\mbox{\scriptsize{c,0}}}}{\Lambda c^2} \right)+\frac{1}{3}\log{\left(\frac{M}{M^\ast}\right)},
\end{equation}
where we have used the definition of $R^\ast$, eq. \eqref{rstar}. Comparing this with eq. \eqref{Idef2}, we see that:
\begin{equation} \label{Iult}
I_{\mbox{\scriptsize{ult}}}=\frac{1}{3}\log \left(\frac{4\pi G \rho_{\mbox{\scriptsize{c,0}}}}{\Lambda c^2} \right)=
\frac{1}{3}\log \left(\frac{4\pi G \rho_{\mbox{\scriptsize{c,0}}}}{ c^2} \right)-\frac{1}{3}\log\Lambda.
\end{equation}
As it is clear from the above equation, $I(a)$ approaches a constant value which is lower for a higher value of $\Lambda$, exactly as it can be inferred from the plot.

An interesting feature of the plot is that at the present epoch ($a=1.00$) $I$ has almost the same value for all models (for all values of $\Omega_{\mbox{\scriptsize{$\Lambda$,0}}}$). Additionally,  $I$ seems to have exactly the same value (the graphs intersect) for every value of $\Omega_{\mbox{\scriptsize{$\Lambda$,0}}}$ a little later, at an point, let us call it $a_{int}$. The inset picture, which focuses around that point, demonstrates that indeed, this is not unique for all graphs. The intersection point a pair of graphs is a little bit different from that of another pair of graphs. However, all are very close, compared to the range where we plot $I$ in the greater figure. For this reason, in the analysis which follows, we thing of an effectively unique point of intersection $a_{int}$ around $a \sim 1.03 - 1.04$ for all graphs. Then we can see that for $a<a_{int}$ we have $I_1(a) > I_2(a)$ if $(\Omega_{\mbox{\scriptsize{$\Lambda$,0}}})_1 >(\Omega_{\mbox{\scriptsize{$\Lambda$,0}}})_2$, while for $a>a_{int}$ we have $I_1(a) < I_2(a)$ if $(\Omega_{\mbox{\scriptsize{$\Lambda$,0}}})_1 >(\Omega_{\mbox{\scriptsize{$\Lambda$,0}}})_2$.

We can understand qualitatively the above behaviour: Consider two models with the same $\Omega_{\mbox{\scriptsize{m,0}}}$ and $(\Omega_{\mbox{\scriptsize{$\Lambda$,0}}})_1>(\Omega_{\mbox{\scriptsize{$\Lambda$,0}}})_2$. A larger value of the cosmological constant in the first model implies that, in order for the two models to end up with the same matter density today, in the past the first model had to have (much) larger 
matter density than the second: $(\Omega_{\mbox{\scriptsize{m,past}}})_1>(\Omega_{\mbox{\scriptsize{m,past}}})_2$, always. In a particular epoch, the dependence of the turnaround overdensity on the value of $\Omega_{\mbox{\scriptsize{$\Lambda$}}}$ is much weaker than the dependence on the value of $\Omega_{\mbox{\scriptsize{m}}}$. And the higher the value of $\Omega_{\mbox{\scriptsize{m}}}$ is, the lower is the value of the turnaround overdensity (it is ``easier" for a structure to turn around).  Thus $(\Omega_{\mbox{\scriptsize{m,past}}})_1 > (\Omega_{\mbox{\scriptsize{m,past}}})_2 \Rightarrow (\delta_{\scriptsize{\mbox{ta,past}}})_1 < (\delta_{\scriptsize{\mbox{ta,past}}})_2$. Then, from the definition of $I(a)$, eq. \eqref{Idef}, $I(a)=\log a -(1/3)\log[(1+\delta_{\scriptsize{\mbox{ta}}}(a))\Omega_{\mbox{\scriptsize{m,0}}} ]$,  a larger value of $\delta_{\scriptsize{\mbox{ta}}}$ leads to a lower (more negative) value for $I$.  Thus, the two models with $(\Omega_{\mbox{\scriptsize{$\Lambda$,0}}})_1>(\Omega_{\mbox{\scriptsize{$\Lambda$,0}}})_2$ had in the past $I_1(a,past)>I_2(a,past)$.

Today, we also have $I_1>I_2$, since for all the time in the past the first model had higher matter density and thus it was easier for turnaround to happen and this continues until today, where the two matter densities are the same. But in the future, the model which has higher $\Lambda$ density today will have lower matter density, and thus then the turnaround overdensity will be higher. So, in the future: $I_1(a, future)<I_2(a, future)$, with ultimate values those predicted by \eqref{Iult}. Since $I_1(a=1) \gtrsim I_2(a=1)$ (they have almost the same value), the point where the two graphs (for $I_1(a)$ and $I_2(a)$) intersect, $a_{int}$ will be very close to the present epoch. Of course, the present is not special. Rather, it seems to be special because we demand $(\Omega_{\mbox{\scriptsize{$m$}}})_1$ for all models to be identical today. Since, not only two, but all models, have similar values for $I$ today and rapidly diverge and spread in the past and in the future, the point of intersection will be almost the same for all graphs, exactly as the plots presented here demonstrate.

\subsection{Constant $\Lambda$ density}

\begin{figure}
\centering
\includegraphics[scale=0.6, clip]{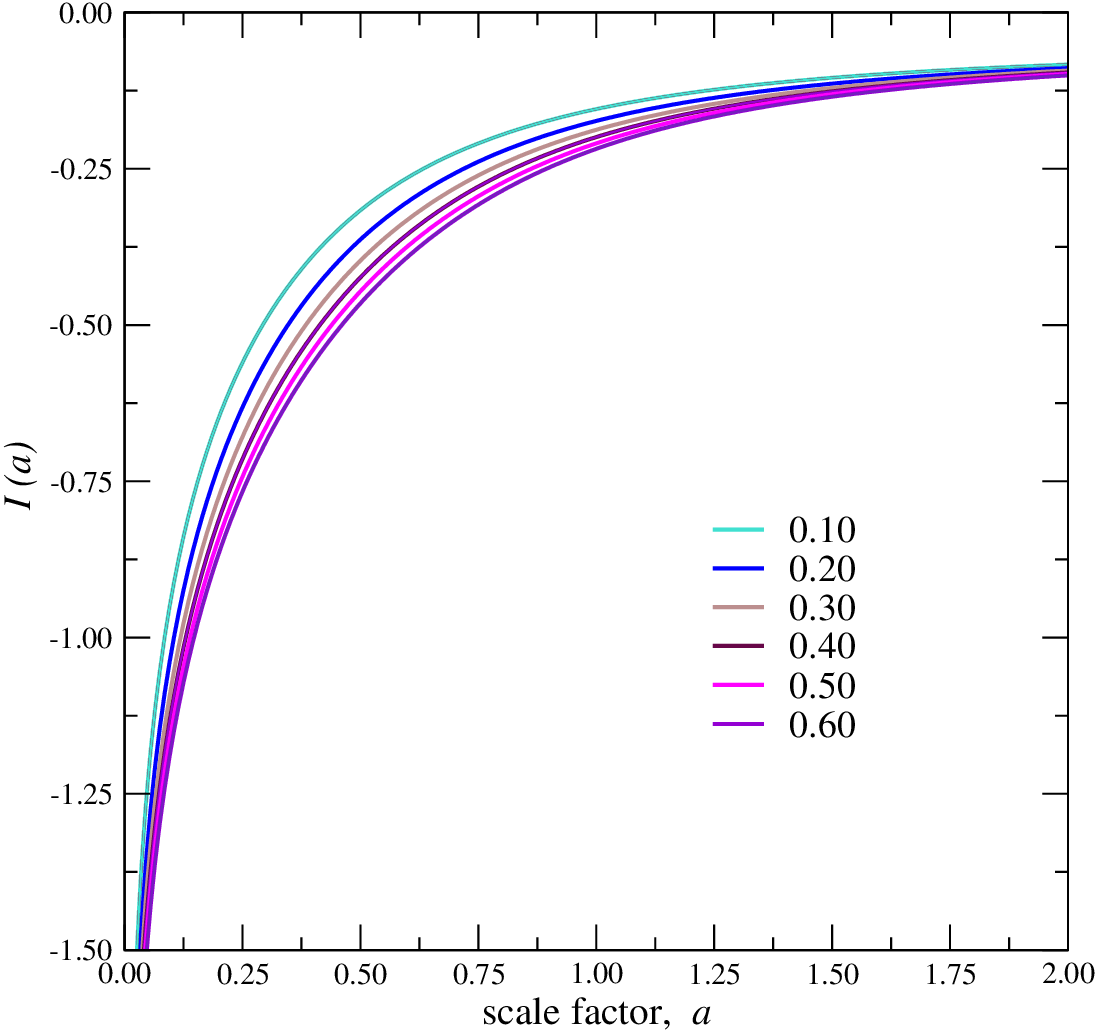} 
\caption{$I(a)$ as a function of $a$, for constant $\Omega_{\mbox{\scriptsize{$\Lambda$,0}}}=0.70$, and different values of $\Omega_{\mbox{\scriptsize{m,0}}}$, in a range from $\Omega_{\mbox{\scriptsize{m,0}}}=0.10$ to $\Omega_{\mbox{\scriptsize{m,0}}}=0.60$.}
\label{fig: figure3}
\end{figure}

It is also interesting to examine the evolution of $I(a)$ for models with constant $\Omega_{\mbox{\scriptsize{$\Lambda$,0}}}$ and different values for the current matter density, $\Omega_{\mbox{\scriptsize{m,0}}}$. We chose $\Omega_{\mbox{\scriptsize{$\Lambda$,0}}}=0.70$, close to the value that is inferred from non-local methods and $\Omega_{\mbox{\scriptsize{m,0}}}$ in the range $[0.10,0.60]$. In figure \ref{fig: figure3} we present $I(a)$ as a function of $a$, for $a \sim 0$ to $a=2.00$ and for the values of cosmological parameters mentioned above.

In figure \ref{fig: figure3} we see that for all models, $I(a)$ reaches the \emph{same} ultimate value. This behaviour can be directly explained referring to eq. \eqref{Iult}: since all models have the same $\Lambda$ density, they will all reach the same ultimate value, $I_{\mbox{\scriptsize{ult}}}$, independently from the value of the current matter density of the Universe.

We see that for two models, with $(\Omega_{\mbox{\scriptsize{m,0}}})_1>(\Omega_{\mbox{\scriptsize{m,0}}})_2$ then $I_1(a)<I_2(a)$ \emph{always}. Since all models have the same $\Lambda$ density, if one model has greater matter density today than the other, it always had and it will always have greater matter density. The turnaround overdensity for the two models is $(\delta_{\mbox{\scriptsize{ta}}})_1<(\delta_{\mbox{\scriptsize{ta}}})_2$. But from the definition of $I(a)$: $I(a)= \log\left(\left[(1+\delta_{\mbox{\scriptsize{ta}}}(a)){\Omega}_{\mbox{\scriptsize{m,0}}} \right]^{-1/3} a \right)$, at a particular epoch the dependence on $\Omega_{\mbox{\scriptsize{m,0}}}$ prevails, thus models with higher matter density give lower values for $I$

\section{Using turnaround for a \emph{local} measurement of $\Lambda$ density} \label{measurement}

\subsection{What do we mean by the term \emph{local}?} 

As we have stated in the introduction, our aim is to find a way to measure locally the value of the cosmological constant. Before presenting how this idea can be implemented using the results presented in sections \ref{Model} and \ref{Dependence}, it is necessary to clarify our use of the term \emph{local}, to avoid misconceptions. 

By the term local, we do not refer to the local universe. What we mean is that each measurement that goes into the method is itself local, i.e. does not depend on what the Universe is doing as a whole (for example, that its expansion is accelerating). It uses an effect of the cosmological constant in relatively  short scales -- the scales of the turnaround radii of cosmic structures. 

Despite of this, our method still has to use data from the high-z universe.  We have to look as far as distant supernovae searches look to find hints of dark energy. This is clear from the results presented in the previous section, where it can be seen that we have to look in the past in order to distinguish between models with the same matter density but different cosmological constant densities (optimally at $a \sim 0.50$, where the difference in the value of $I$ for different models becomes maximum).

\subsection{A worked example: towards a local proof that $\Lambda \neq 0$}

Suppose that we are able to measure turnaround radii, $R_{\mbox{\scriptsize{ta}}}$, and masses, M, of structures at an epoch $a=0.50$ $(z=1.00)$, where the difference of the predicted values of $I$ for different models is maximum, with fractional uncertainties:
\begin{equation} \label{frfm}
f_R \equiv \frac{\sigma_R}{R_{\mbox{\scriptsize{ta}}}(a)} \sim 0.05, \quad f_M \equiv \frac{\sigma_M}{M} \sim 0.3,
\end{equation}
with $\sigma_R$ and $\sigma_M$ the uncertainties in radius and mass, respectively; i.e., supposing that we will be able to achieve at that epoch similar uncertainties, as those achieved for the local universe using current measurement techniques (see the relevant discussion in \cite{PavTom} and the references therein). 

\begin{figure}
\centering
\includegraphics[scale=0.6, clip]{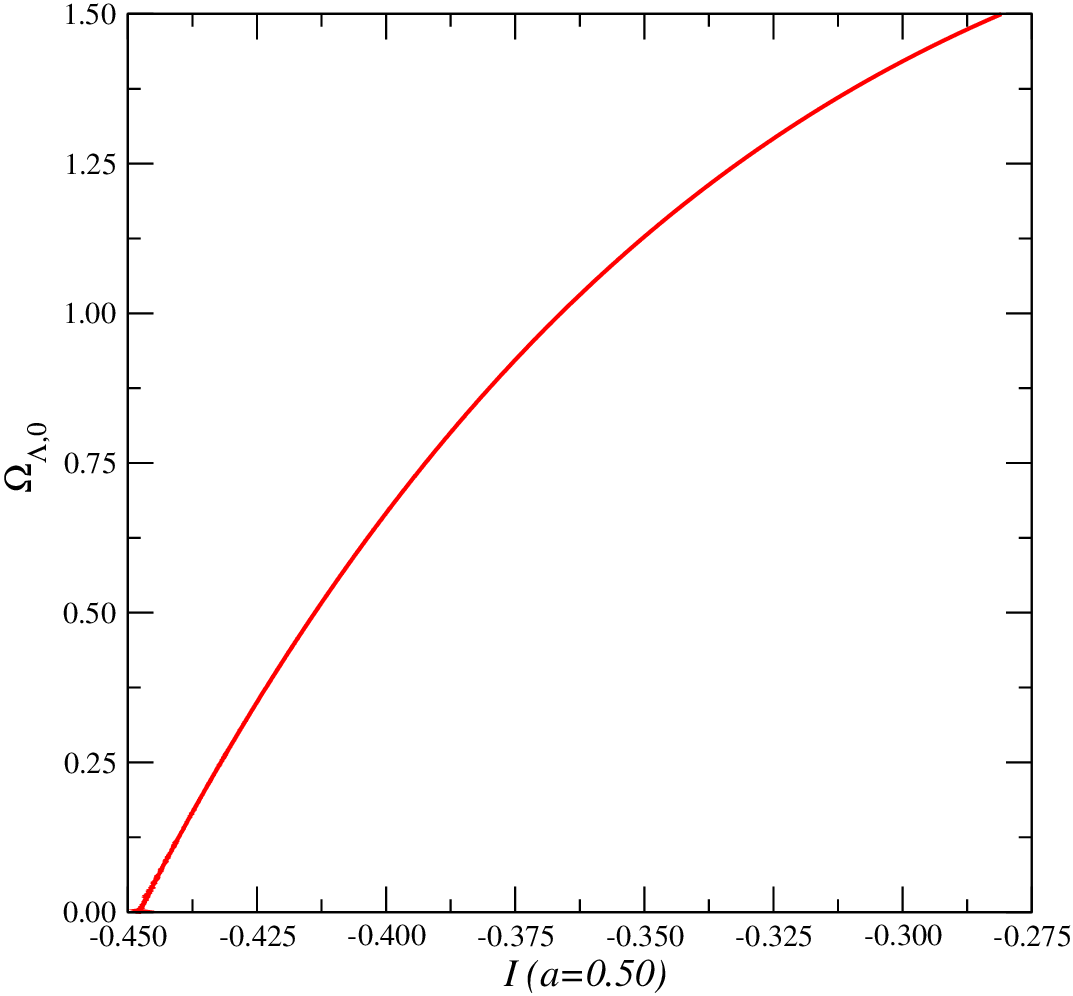} 
\caption{ $\Omega_{\mbox{\scriptsize{$\Lambda$,0}}}$ as a function of $I$, at $a=0.50$ and $\Omega_{\mbox{\scriptsize{m,0}}}=0.30$, constant.}
\label{fig: figure5}
\end{figure}

As an example of the sensitivity of our method in determining the value of the cosmological constant, we consider the following question: how many structures we have to examine in order to establish with confidence that $\Lambda \neq 0$? 
Here we work in a model with $\Omega_{\mbox{\scriptsize{m,0}}}=0.30$. In applying this method, trying to prove locally the existence of $\Lambda$, we can use results of other methods of getting the matter density of the Universe (BAO's, galaxy clusters, e.g., \cite{BAOS1,BAOS2,Eke}) since they are also --in the sense described in the previous sub-section-- local methods and do not harm the locality of the test. Of course, measuring the matter density of the Universe using our method is also possible, since the difference in matter density affects the evolution of $I(a)$, but this would make much more complicated the practical implementation of the test.
 
 In what follows, we consider that $a$ (through $z$) can be measured with much higher accuracy than masses and radii of structures, so it is not a source of error in our calculations  and it is treated just as a parameter. 
By measuring the turnaround radius and the mass of a structure, a value of $I(a)$ can be obtained (by rearranging eq. \eqref{Idef2}):
\begin{equation}
I(a) = \log\left(\frac{R_{\mbox{\scriptsize{ta}}}(a)}{R^\ast}\right)- \frac{1}{3}\log\left(\frac{M}{M^\ast}\right).
\end{equation}
Errors in the measurements of radius and mass result in an uncertainty, $\sigma_I$, in the calculated value of $I$. Error propagation gives \cite{Error}:
\begin{equation}
\sigma_I^2=\frac{1}{\ln^210} \left(\frac{1}{9}\frac{\sigma_M^2}{M^2}+\frac{\sigma_R^2}{R^2_{\mbox{\scriptsize{ta}}}(a)} \right) \Rightarrow \sigma_I = \frac{1}{\ln10} \left(\frac{1}{9}f_M^2+f_R^2 \right)^{1/2},
\end{equation}
where we have used the definitions of $f_M$ and $f_R$.
Plugging the values for $f_M$ and $f_R$ presented in eq. \eqref{frfm}, we get:
\begin{equation} \label{sigmas}
\sigma_I \sim 0.05.
\end{equation}

Let us now imagine that we perform $N$ measurements of masses and radii of $N$ different structures (for simplicity, consider that we measure all $N$ structure exactly at $a=0.50$) and we get $N$ different values for $I$. Denote this mean value $\langle I \rangle $. If all values have the same error $\sigma_I$, (all measurements have the same fractional errors $f_R$ and $f_M$), then the error of the mean, $\sigma_{\langle I \rangle}$, will be \cite{Error}
\begin{equation}
\sigma_{\langle I \rangle}= \frac{\sigma_I}{\sqrt{N}}.
\end{equation}
From the above equation we can find the number of measurements we have to perform in order to get a particular error in the mean $\langle I \rangle$, $\sigma_{\langle I \rangle}$, assuming that the error of every individual value of $I$ is the same and equal to $\sigma_I$:
\begin{equation} \label{nu}
N = \left(\frac{\sigma_I}{\sigma_{\langle I \rangle}} \right)^2
\end{equation}

Figure \ref{fig: figure5} presents the inferred value for $\Omega_{\mbox{\scriptsize{$\Lambda$,0}}}$ for any measured value of $I$ at an epoch $a=0.50$. Suppose now that we measure a mean value of $I$ about $-0.400$ which corresponds to a value for $\Omega_{\mbox{\scriptsize{$\Lambda$,0}}} \sim  0.68$, the value inferred from other techniques. To establish a $5 \sigma$ confidence that $\Omega_{\mbox{\scriptsize{$\Lambda$,0}}} \neq 0$, we demand this observed mean value of $I$ to be five standard deviations away from the value of $I$ which corresponds to zero  value for the cosmological constant, which is $I \sim -0.450$ (from fig. \ref{fig: figure5}). This allows us to calculate the value of the error in the mean that establishes the desired confidence:

\begin{equation}
5 \sigma_{\langle I \rangle} = -0.400 - (-0.450) \Rightarrow \sigma_{\langle I \rangle} \sim 0.01
\end{equation}
If all measurements of $I$ have the same error, and equal to that presented in \eqref{sigmas}, then we can determine the number of measurements, $N$, we have to make to get the necessary accuracy in $\langle I \rangle$, from \eqref{nu} :
\begin{equation}
N \sim \left(\frac{0.05}{0.01}\right)^2 = 25
\end{equation}

The above discussion, even with the simplifying assumptions that all structures are exactly at $a=0.50$ and have the same uncertainties in mass and radius for all structures (also, extrapolating the current measurement techniques for the local universe to the universe at $a \sim 0.50$), gives a rough -- order-of-magnitude -- estimation of the number of measurements that have to be done in order to confidently establish a local proof that a non-zero cosmological constant exists.

\section{Discussion} \label{Discussion}

Based on the need for the cosmological constant to be weak enough to allow gravitational bound structures to form, anthropic arguments have long before used to set upper bounds for its value (e.g., \cite{Weinberg}). Detailed studies concerning galaxy formation put tighter upper bounds for the value of $\Omega_{\mbox{\scriptsize{$\Lambda$,0}}}$, turned out to be very close to the value inferred later from the measured accelerated expansion of the Universe \cite{Galform}.

In the present work, we used structure formation to demonstrate a way not only to set a local upper bound for the cosmological constant but rather to show that a  local \emph{measurement} of $\Omega_{\mbox{\scriptsize{$\Lambda$,0}}}$ is in principle possible. Especially that it is easy to use the evolution of the turnaround radius of cosmic structures in order to set a \emph{lower} bound to the value of $\Omega_{\mbox{\scriptsize{$\Lambda$,0}}}$ -- in other words, to obtain a local proof that the cosmological constant has a non-vanishing value. The merits of such a proof (or disproof) have been discussed in the introduction.

In our work we have mainly been focused on how to use the evolution of the turnaround radius to measure the cosmological constant, not the matter density of the Universe. However, measuring the matter density using this method it is also possible, in principle. We have seen that today the value of the turnaround radius (or the function $I(a)$ defined in the text, today) has a extremely weak dependence on $\Omega_{\mbox{\scriptsize{$\Lambda$,0}}}$ but depends on the value of $\Omega_{\mbox{\scriptsize{m,0}}}$. Thus, using observations from the very close Universe we can determine the value of the matter density and then use it as an input, in order to measure the cosmological constant by going backwards in time, i.e. to the distant Universe. Although it is important that this method can be used to measure both parameters, in practice this would be quite complex. If our target is a local measurement of the cosmological constant, we can use as input for the value of matter density of the Universe the value obtained using other methods, which are also local.

Our analysis is based on the spherical collapse model, which is a simple model for the description of structure formation. However, we expect the results presented here to hold at least qualitatively. In \cite{PavTom} and \cite{myRoy} it is discussed why the effect of non-sphericities is not expected to be severe at turnaround scales.   Since our aim was to give a proof-of-principle about the ability to use the evolution of the turnaround radius in order to measure cosmological parameters  --especially the value of the cosmological constant-- the treatment presented here is adequate. The implementation of this idea is not easy, and before using actual observations, in order to get accurate values for $\Omega_{\mbox{\scriptsize{m,0}}}$ and $\Omega_{\mbox{\scriptsize{$\Lambda$,0}}}$,  the evolution of $I$ has to be benchmarked at higher accuracy using numerical simulations.

\acknowledgments
 Research implemented under the “ARISTEIA II” Action of the
 Operational Program “Education and Lifelong Learning” and is
 co-funded by the European Social Fund (ESF) and Greek National
 Resources. The work of TNT was partially supported by the European
 Union Seventh Framework Program (FP7-REGPOT-2012-2013-1) under grant
 agreement No. 316165. VP and DT acknowledge support by the European Commission Seventh Framework Programme through grants PCIG10-GA-2011-304001 ``JetPop'' and PIRSES-GA-2012-31578 ``EuroCal''.



\end{document}